\documentclass[conference]{IEEEtran}

\usepackage{comment}
\IEEEoverridecommandlockouts
\usepackage{cite}
\usepackage{amsmath,amssymb,amsfonts}
\usepackage{algorithmic}
\usepackage[linesnumbered,ruled,vlined]{algorithm2e}
\usepackage{graphicx}
\usepackage{textcomp}
\usepackage{xcolor}
\usepackage[T1]{fontenc}
\usepackage{subfigure}
\usepackage{multirow}
\usepackage{booktabs}
\usepackage{enumitem}
\usepackage{pifont}
\usepackage{graphicx}
\usepackage{wrapfig}

\def\BibTeX{{\rm B\kern-.05em{\sc i\kern-.025em b}\kern-.08em
    T\kern-.1667em\lower.7ex\hbox{E}\kern-.125emX}}

\usepackage[bookmarks=true,breaklinks=true,colorlinks,linkcolor=blue,citecolor=blue,urlcolor=blue]{hyperref}

\usepackage{comment}
\usepackage{xspace}
\newcommand{\designTitle}{QuantumLeak\xspace}
\newcommand{\design}{\textit{QuantumLeak}\xspace}

\begin{document}
\title{\designTitle: Stealing Quantum Neural Networks from Cloud-based NISQ Machines \vspace{-4pt}
\thanks{This paper is accepted by International Joint Conference on Neural Networks (IJCNN) 2024.}}
% \author{\IEEEauthorblockN{Anonymous Authors}}

\author{
\begin{tabular}{cccccc}
Zhenxiao Fu$^*$ & Min Yang$^*$ & Cheng Chu$^*$ & Yilun Xu$^\dag$ & Gang Huang$^\dag$ & Fan Chen$^*$ \\
\multicolumn{3}{c}{$^*$Indiana University Bloomington} & \multicolumn{3}{c}{$^\dag$Lawrence Berkeley National Laboratory}\\
\multicolumn{3}{c}{$^*$\{zhfu, my36, chu6, fc7\}@iu.edu} & \multicolumn{3}{c}{$^\dag$\{yilunxu, ghuang\}@lbl.gov}\\
\end{tabular}
\vspace{-18pt}
}

\maketitle

\begin{abstract}
Variational quantum circuits (VQCs) have become a powerful tool for implementing Quantum Neural Networks (QNNs), addressing a wide range of complex problems. Well-trained VQCs serve as valuable intellectual assets hosted on cloud-based Noisy Intermediate Scale Quantum (NISQ) computers, making them susceptible to malicious VQC stealing attacks. However, traditional model extraction techniques designed for classical machine learning models encounter challenges when applied to NISQ computers due to significant noise in current devices. In this paper, we introduce \design, an effective and accurate QNN model extraction technique from cloud-based NISQ machines. Compared to existing classical model stealing techniques, \design improves local VQC accuracy by 4.99\%$\sim$7.35\% across diverse datasets and VQC architectures. 
% Finally, we discuss potential defense mechanisms against \design attack.

\end{abstract}


\IEEEpeerreviewmaketitle

\section{Introduction}

Quantum Neural Networks (QNNs) have shown remarkable capabilities across diverse domains, including natural language processing natural language processing~\cite{chen2020variational}, object recognition~\cite{Chu:ISLPED2022}, and financial analysis~\cite{Egger:TQE2020}. At their core are Variational Quantum Circuits (VQCs)~\cite{Mitarai:physA2018}, designed with specific quantum ansatz and trained on targeted datasets. Developing an accurate VQC for a task demands substantial domain expertise and costly data collection. Consequently, VQCs are valuable intellectual properties (IPs) that warrant robust protection. Typically offered as cloud services on Noisy Intermediate Scale Quantum (NISQ) machines, VQCs have attracted adversaries seeking to steal them. Through various \textit{model extraction} attacks~\cite{tramer2016stealing, papernot:2017practical, Jacson:IJCNN2018, Yu:NDSS2020}, an adversary can create a local substitute VQC mimicking the functionality of the victim VQC on a cloud-based NISQ machine by querying it multiple times, posing a severe threat to the VQC owner's IP.

However, the model extraction techniques previously proposed~\cite{tramer2016stealing, papernot:2017practical, Jacson:IJCNN2018, Yu:NDSS2020} were initially designed for pilfering classical machine learning models. Our preliminary results indicate that the na\"ive application of these schemes yields only suboptimal levels of accuracy when employed for VQC extraction from NISQ servers. This decrease in accuracy can be primarily attributed to the substantial inherent quantum noise pervasive in NISQ computers~\cite{Clerk:RMP2010,sarovar2020detecting,Huo:NJP2017,Lax:PR1966}. NISQ devices are vulnerable to various forms of quantum noise originating from environmental interactions~\cite{Clerk:RMP2010}, qubit crosstalk~\cite{sarovar2020detecting}, imperfect controls~\cite{Huo:NJP2017}, and leakage~\cite{Lax:PR1966}. Moreover, different NISQ technologies may encounter distinct levels of noise due to their diverse architectures and control mechanisms~\cite{Clerk:RMP2010}, as well as their interactions with varying environmental conditions. 

\vspace{1pt}
\noindent\textbf{{Contributions}}.
In this paper, we propose \design, a novel quantum model extraction attack utilizing an ensemble of local QNN learners to enhance successful extraction from a NISQ machine, even in the presence of quantum noise. The contributions of this paper are summarized as follows:

\begin{itemize}[leftmargin=*, nosep, topsep=0pt, partopsep=0pt]
\item \textbf{Ensemble-based QNN Extraction Attack}.
  To the best of our knowledge, this work marks the first QNN extraction attempts. We investigate state-of-the-art (SOTA)  extraction attacks in the context of QNN theft from NISQ devices, uncovering their inferior performance.
  We propose~{\textit{\design}}, which strategically leverages an ensemble of local substitute models to improve extraction attack success. \design~proceeds through three major phases: victim QNN query and data bagging, substitute QNN initialization and training, and decision fusion. We provide comprehensive insights into each phase, discussing techniques dedicated to enhancing effectiveness throughout the process.  

\item \textbf{Robust QNN Learning with Noisy Data}.
  To enhance the robustness of local substitute QNN training with queried noisy data, we systematically investigate various training optimization techniques within the proposed~\design~framework. This exploration leads to the identification of a generalized formulation of the Huber loss, a key enhancement that substantially improves both the robustness and accuracy of the learning process.

\item \textbf{Demonstrated Success in QNN Extraction}.
  We conducted a comprehensive evaluation of~\design~across diverse VQC architectures and ensemble configurations. Experimentation on IBM NISQ devices showed that~\design~achieves an improvement of 4.99\%$\sim$7.35\% in the accuracy of the locally-generated QNN compared to CloudLeak~\cite{Yu:NDSS2020}, a SOTA classical model extraction attacks. Additionally, we discuss potential countermeasures to defend against our~\design~QNN extraction attacks.  
\end{itemize}

\noindent\textbf{{Limitations and Scope}}.
The selected benchmark dataset (e.g., MNIST and Fashion-MNIST) and the implementation scale (e.g., up to 27 qubits) in this paper are constrained by existing classical-to-quantum encoding inefficiencies, our limited access to IBM quantum computers, and high noise rates on NISQ computers.
As technological advancements progress, notable improvements in the implementation scale and performance of~\design~are anticipated.
This work initiates research on QNN model extraction attacks on NISQ computers, paving the way for future investigations in this domain and the broader area of quantum security.

\begin{figure}[t!]
\centering
\includegraphics[width=\linewidth]{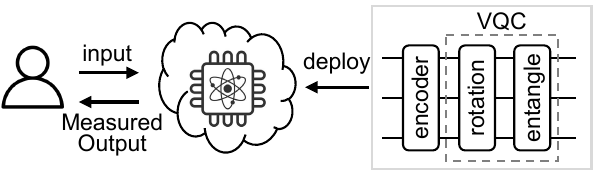}
\vspace{-0.3in}
\caption{The QNN-as-a-service on a NISQ computer.}
\vspace{-0.24in}
\label{f:background_qnn}
\end{figure}

\vspace{-4pt}
\section{Background and Problem Definition}
\vspace{-1pt}

\subsection{Key Concepts in Quantum Neural Networks}

\noindent\textbf{Basics of QNNs}.
A \textit{qubit} is described by a linear combination in a two-dimensional state space, denoted as $|\phi\rangle=\alpha|0\rangle+\beta|1\rangle$, where $|0\rangle$ and $|1\rangle$ form an orthonormal basis, and $\alpha, \beta \in \mathbb{C}$ with $|\alpha|^2+|\beta|^2=1$.
A \textit{quantum gate} on an $n$-qubit state defines a $2^n\times2^n$ unitary matrix (i.e., $\texttt{U}$) that transforms the input state $|\phi\rangle$ to $|\psi\rangle=\texttt{U}|\phi\rangle$.
NISQ computers feature a restricted universal native gate set, comprising only a few types of 1-Qubit gates and typically a single type of 2-Qubit gate. For instance, IBM quantum computers exclusively support two 1-Qubit gates (e.g., \texttt{U2}, \texttt{U3}) and one 2-Qubit gate (e.g., \texttt{CNOT}).
A \textit{quantum algorithm} is represented as a circuit, composed of a sequence of quantum gates operating on a set of suitably initialized qubits. 
As a representative NISQ algorithm, the core of a QNN involves VQCs\cite{Mitarai:physA2018, Egger:TQE2020} that are designed with a specific quantum ansatz and trained on domain-specific datasets. As conceptually illustrated in Figure~\ref{f:background_qnn}, a QNN consists of three key components: a data encoder, $\mathbf{E(x)}$; a multi-layer trainable circuit, $U(\mathbf{\theta})$; and a quantum measurement layer, $\mathbf{M}$. These components collectively form the function $\mathbf{Q} = \mathbf{M} \circ U(\mathbf{\theta}) \circ \mathbf{E(x)}$. 
The QNN generates the final predicted label by applying a softmax operation on the raw probability output vector. Similar to classical neural networks, parameters in a VQC can be trained using a classical optimizer\cite{Zhu:SCIENCE2019}. 
\textit{The creation of an accurate VQC demands profound domain expertise and entails expensive data acquisition processes, underscoring the valuable intellectual property (IP) nature of VQCs and the need for robust protection measures.}

\vspace{2pt}\noindent\textbf{Quantum Cloud Computing}. 
The high costs associated with quantum computers and the intricacies of designing precise quantum algorithms pose significant barriers for individual users. Consequently, average users typically leverage NISQ hardware through a quantum cloud server, either by downloading free services or purchasing QNNs from specific quantum algorithm providers via QNN-as-a-Service (QNNaaS).
In this model, demonstrated in Figure~\ref{f:background_qnn}, users interact with QNNaaS by submitting queries with their local data. The server then executes the QNN on its NISQ devices, measures the outputs, and provides classical probability results to end-users for subsequent post-processing, ultimately yielding a classical prediction. The restricted privilege levels inherent in QNNaaS limit users from gaining insights into crucial aspects such as the QNN architecture, training datasets employed by the cloud, and additional hardware characteristics, including error rates and power traces.
\textit{This lack of transparency presents challenges for adversaries seeking to extract QNN models and general information from the cloud.}

\begin{table}[t]
\centering
\caption{Error rates in current NISQ quantum computers.}
\label{tab:quantum_noise}  
\vspace{-10pt}
\begin{tabular}{|l||l|l|l|l|}		
\toprule
\multirow{2}{*}{\textbf{NISQ Device}} &\multirow{2}{*}{\textbf{Qubit \#}} &\multicolumn{3}{c|}{\textbf{Error Rate}} \\\cline{3-5}
& & \textbf{SPAM} & \textbf{1Q-Gate} & \textbf{2Q-Gate} \\ \hline\midrule
IonQ\_Forte  &32 &0.50\%  &0.020\% &0.40\%  \\ \hline
IBM\_Auckland & 27 &0.34\% &0.197\% &2.44\%   \\\hline
IBM\_Kolkata  & 27 &0.35\% &0.177\% &2.87\%   \\\bottomrule
\end{tabular}
\vspace{-0.26in}
\end{table} 

\vspace{2pt}\noindent\textbf{Quantum Noises}.
Noise is an inherent characteristic and a significant limitation in most current NISQ devices, stemming from imperfections in fabrication, control, and state preparation/measurement (SPAM), which manifest as observable errors. These errors primarily include SPAM errors~\cite{ryan2021realization}, gate errors~\cite{nielsen2002quantum, rudinger2019probing}, and circuit-level crosstalk errors~\cite{ahsan2022quantum}.
Table~\ref{tab:quantum_noise} presents error rates for SPAM and gate errors in two prominent NISQ technologies: trapped-ion (e.g., \texttt{IonQ\_Forte}) and superconducting (e.g., \texttt{IBM\_Auckland} and \texttt{IBM\_Kolkata}). Note that crosstalk errors, which vary depending on the specific quantum circuit and its physical mapping to a target NISQ computer, are not included in the table. Subsequently, we provide a brief discussion on the characteristics of each type of noise.

\vspace{2pt}
\begin{itemize}[leftmargin=*, topsep=0pt, partopsep=0pt, itemsep=0pt]
\item \textit{SPAM errors}.
The initialization of a quantum state and the subsequent measurement of its state introduces errors~\cite{ryan2021realization}. In the context of QNNs, qubits are typically initialized in the ground state, making measurement errors a dominant form of SPAM error. For example, reading $|1\rangle$ while the qubit is in the $|0\rangle$ state, and vice versa. Measurement errors are presently modeled using a 2$\times$2 probability matrix and can be effectively mitigated through readout-error correction and/or post-measurement adjustment~\cite{nachman2020unfolding}.
\item \textit{Gate errors}. 
Quantum gates suffer from errors, categorized into two types: \textit{coherent} and \textit{incoherent} errors. Managing \textit{incoherent errors} is generally more straightforward, as they can often be effectively modeled as depolarizing noise~\cite{nielsen2002quantum}. On the other hand, \textit{coherent errors} typically stem from miscalibrations in control parameters, resulting in similar or even drifting~\cite{rudinger2019probing} during consecutive executions of a quantum circuit. This phenomenon introduces a systematic bias in the final output, making \textit{coherent errors} more challenging to address. As numerically demonstrated in Table~\ref{tab:quantum_noise}, 2-Qubit gates dominate gate errors, exhibiting error rates orders of magnitude higher than those observed in 1-Qubit gates.
\item \textit{Crosstalk errors}.
Larger NISQ systems (i.e., >10 qubits) exhibit crosstalk noise, causing visible deviations from ideal behavior. Crosstalk characteristics depend on the quantum circuit architecture and vary significantly across physical platforms with different technologies. Recent research~\cite{ahsan2022quantum} highlighted a 20\% increase in phase-flip error and a 33\% decrease in gate fidelity in a circuit involving only 9 \texttt{CNOT} gates on the \texttt{IBM\_Melbourne} computer~\cite{ahsan2022quantum}.
\end{itemize}
\textit{Despite recent advancements in noise-robust QNN designs~\cite{Sukin2019_vqcc14, chen2020variational, Chu:ISLPED2022}, several error types, including coherence errors and crosstalk errors, persist unaddressed in these schemes. Consequently, the significant quantum noise diminishes QNN accuracy when executed on real devices.}

\subsection{Problem Definition}
\noindent\textbf{Problem Formulation}.
An adversary aims to accurately construct a local substitute QNN, denoted as $Q_A$, that closely mirrors the victim QNN $Q_V$ deployed in the cloud, by querying the victim QNN with unlabeled inputs and collecting the corresponding responses. The model extraction process involves the following steps:
(1) \textit{Initiate substitute QNN}: The adversary initiates a basic substitute local QNN model, $Q_A$, from a function family $\mathbb{Q}$ and sets its hyperparameters.
(2) \textit{Victim QNN query}: The adversary gathers an initial set of unlabeled samples, denoted as $S$, by querying the victim model $Q_V$ and recording the responses in a dataset $D=\{S, Q_V(S)\}$.
(3) \textit{Substitute model training}: The adversary trains the substitution QNN $Q_A$ with the dataset $D$ using a specific training strategy, optimizer, and decision-fusion scheme.

\vspace{2pt}\noindent\textbf{Threat Model}.
In the QNNaaS paradigm, users provide input data to the NISQ machine, which utilizes the QNN to produce raw probability output vectors. Since the NISQ machine lacks classical computing capability, it directly returns the raw probability output vectors to the client, who can obtain the final predicted labels by performing classical softmax operations on the raw probability output vectors. For economic and privacy reasons, the specific details of the QNN remain undisclosed. In this paper, we assume that the victim QNN is treated as a black box by the adversary, lacking access to internal information like circuit architecture, training dataset, hyperparameters, and quantum gate parameters. Moreover, the adversary cannot measure the noises on the NISQ machine. The adversary's capabilities are restricted to submitting inputs and receiving corresponding raw probability output vectors. By collecting input-output pairs, the adversary can train a local substitute QNN that closely emulates the performance of the cloud-based QNN within the black-box setting. This substitute QNN empowers the adversary to perform an unlimited number of query tasks without incurring any costs.

\section{Related Work and Motivation}

\subsection{Related Work}\vspace{-2pt}
\noindent\textbf{Model Extraction Attacks}.
The first classical model extraction attack~\cite{tramer2016stealing} exclusively targeted simple models like logistic regression, SVM, decision trees, and shallow neural networks. Subsequent research, with notable contributions as seen in~\cite{papernot:2017practical}, demonstrated that adversarial examples from a substitute model can lead to a significant misclassification rate in a black-box model. This concept was further expanded upon in~\cite{Jacson:IJCNN2018}, which relaxed black-box setting requirements to enable the learning of useful information from data samples different from the original training set. The most recent development, CloudLeak~\cite{Yu:NDSS2020}, employs an active learning mechanism to substantially reduce query number budget, enhancing the stealthiness of the model extraction process.

\vspace{2pt}\noindent\textbf{Ensemble Learning}.
Ensemble learning~\cite{hansen:1990neural}, constructs a predictive model by combining the strengths of a collection of simpler base models, aiming to minimize the expected prediction error. Key methods in ensemble construction~\cite{ferreira2012boosting} include boosting, randomization, and bagging. Boosting involves leveraging a committee of weak learners that evolves over time. Randomization methods consist of estimating the single base model with a randomly perturbed training algorithm. Bagging fits the same model to different training sets, creating a committee of independent weak learners, and the ensemble prediction is then obtained through decision fusion. Bagging, particularly effective in the presence of noisy data, enhances accuracy by mitigating variance components.

% \vspace{2pt}\noindent\textbf{Learning with Noisy Labels}.
% Most existing methods for training DNNs with noisy labels seek to correct the loss function. The correction can be categorized in two types. The first type treats all samples equally and correct loss either explicitly or implicitly through relabeling the noisy samples. For relabeling methods, the noisy samples are modeled with directed graphical models (Xiao et al., 2015), Conditional Random Fields (Vahdat, 2017), knowledge graph (Li et al., 2017b), or DNNs (Veit et al., 2017; Lee et al., 2018).  Most existing methods for training DNNs with noisy labels seek to correct the loss function. The correction can be categorized in two types. The first type treats all samples equally and correct loss either explicitly or implicitly through relabeling the noisy. 

\begin{figure}[t!]
%\vspace{-0.1in}
\centering
\includegraphics[width=\linewidth]{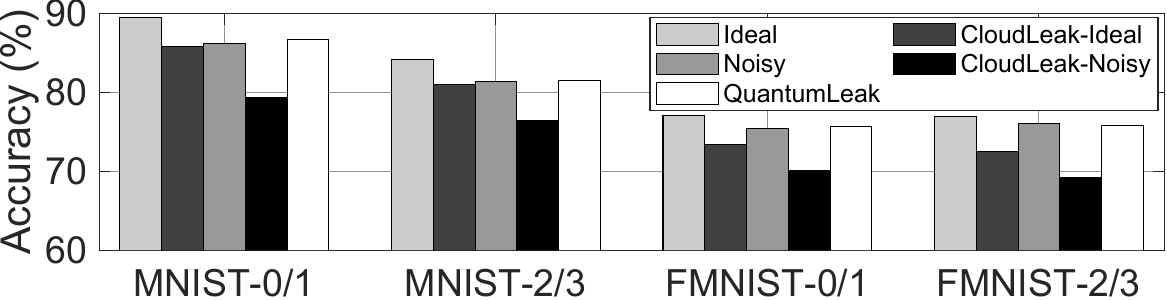}
\vspace{-0.3in}
\caption{The accuracy of victim QNN and CloudLeak attacks~\cite{Yu:NDSS2020}.}
\label{f:motivation}
\vspace{-0.26in}
\end{figure}

\begin{figure*}[t!]
\centering
\includegraphics[width=\linewidth]{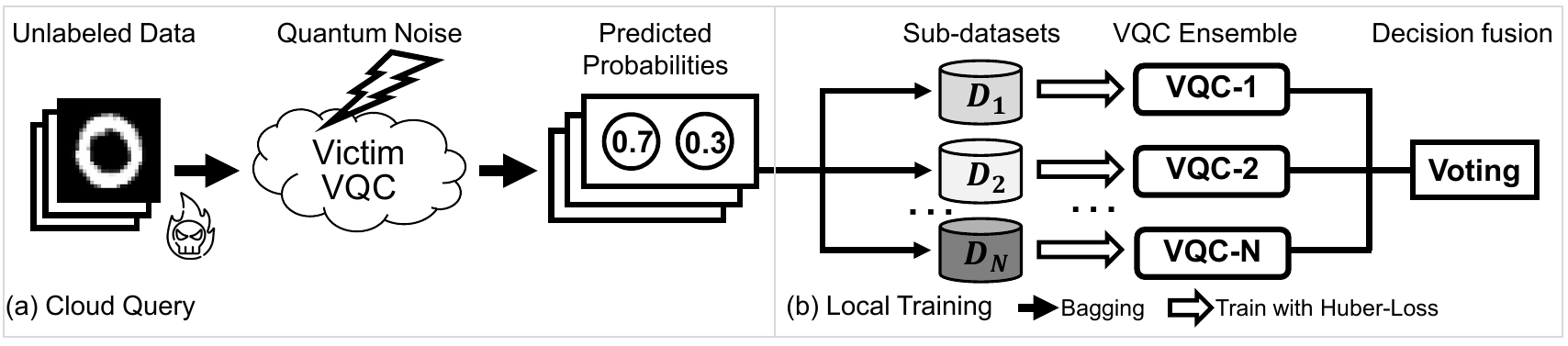}
\vspace{-0.28in}
\caption{The overall architecture and working flow of \designTitle.}
\label{f:overall}
\vspace{-0.2in}
\end{figure*}

\vspace{-2pt}\subsection{Motivation}\vspace{-2pt}
\noindent\textbf{Reduced QNN Accuracy on NISQ Computers}.
We utilize a SOTA QNN, \textit{Circuit 14} in~\cite{Sukin2019_vqcc14}, to train a victim model. This model comprises an amplitude encoder and a 2-layer VQC, and is trained on the MNIST and Fashion-MNIST datasets. The inference accuracy in simulation (i.e., \textit{Ideal}) and on \texttt{IBM\_Auckland} (i.e., \textit{Noisy}) with the respective datasets, is illustrated in Figure~\ref{f:motivation}.
The notable accuracy reduction on the NISQ computer suggests that queries to the victim QNN from NISQ machines yield noisy labels with high levels of inaccuracy. This phenomenon impedes the training of an accurate substitute model, highlighting the challenges associated with using data generated from NISQ devices.

\vspace{2pt}\noindent\textbf{Ineffectiveness of Existing Techniques}.
Operating under a strong white-box assumption, we implemented a local substitute QNN identical to the victim QNN and employed the same datasets as the cloud model. We applied the SOTA technique, CloudLeak~\cite{Yu:NDSS2020}, to extract the QNN through query respectively the \textit{Ideal} and \textit{Noisy} QNNaaS indicated in Figure~\ref{f:motivation}. 
Advanced techniques within CloudLeak, including transfer learning and adversarial samples, were incorporated to enhance the accuracy of the local QNN model. 
More specifically, transfer learning involved pre-training the local QNN model with 3,000 images from publicly available datasets, ensuring similar data distributions to the victim QNN~\cite{Yu:NDSS2020}. Adversarial samples, crucial for determining the separating surface of the victim QNN, were constructed using the methodology proposed in~\cite{Yu:NDSS2020}. The adversary initiated an unlabeled dataset with 6000 images within the problem domain and 500 images outside the problem domain, submitting it to the victim QNN for queries. The server returned the raw probability vectors of the dataset, enabling the adversary to train a local QNN using the labeled dataset. 
As illustrated in Figure~\ref{f:motivation}, although CloudLeak extract a substitute QNN with acceptable accuracy from the  \textit{Ideal} server, the accuracy in the \textit{Noisy} setting shows a significant up to 10.15\% reduction in accuracy from the victim QNN.
\textit{The preliminary study results have motivated us to develop a QNN extraction attack scheme capable of learning an accurate local substitute QNN model using noisy labels obtained from queries to a NISQ server.}

\section{\designTitle}\vspace{-4pt}

\noindent\textbf{Overview}.
Figure~\ref{f:overall} illustrates the overall architecture and working process of~\design, encompassing three main phases: initializing substitute QNN at the local, querying the victim QNN on the cloud, and training the local substitute model.

\vspace{2pt}
\begin{itemize}[leftmargin=*, topsep=0pt, partopsep=0pt, itemsep=0pt]
\item \textbf{Initiating Substitute Model}.
Unlike classical machine learning, QNNs currently adopt a standardized structure, primarily utilizing a VQC that consists of a rotation layer with 1-Qubit gates and entanglement layers employing 2-Qubit gates arranged in a ring structure. In our investigation, we consider mainstream QNN models~\cite{Sukin2019_vqcc14, chen2020variational, Chu:ISLPED2022} as the zoo of candidate QNN models. From this selection, one model is chosen to serve as the primitive substitute local QNN. We initialize the variational gate parameters following a Gaussian distribution.

\item \textbf{Querying the Victim QNN}.
The victim QNN model is queried through the QNNaaS interface, obtaining predictions to form the base dataset. Our objective is to enhance the model accuracy through ensembles of local QNNs, capitalizing on their collective decision-making capabilities. The robustness afforded by multiple models surpasses the predictive power of a single QNN, particularly in the presence of noisy labels. To achieve this, an iterative bagging approach is applied to the dataset, facilitating the training of each ensemble committee QNN individually with a subset. 

\item \textbf{Training the Local Substitute Model}.
In QNN design, the topology of the VQC ansatz and the number of repeated VQC layers play a crucial role in determining the final model accuracy. We discuss these configuration choices within the~\design~framework. Moreover, in the context of model extraction attacks, minimizing queries to the victim model is ideal. This not only reduces query costs but also mitigates the risk of detection. For adversaries, querying the QNNaaS should be more cost-effective than independently training a model with similar functionality. However, a high volume of queries may raise concerns with the cloud service provider, potentially leading to account restrictions or other countermeasures. We delve into a discussion and present experimental results highlighting the impact of query budget. Leveraging the enhanced learning capability of ensembles, our results demonstrate that~\design~outperforms existing approaches with comparable query budgets, achieving superior accuracy.
\end{itemize}

\subsection{Initiating Substitute Model}
The exacting topological constraints imposed by mainstream QNNs limit the exploration space within the QNN model zoo. Even in the context of assuming a black-box attack, wherein the adversary lacks knowledge of the victim QNN's internal structure, there exists a substantial probability that the adversary might inadvertently incorporate the victim QNN architecture into their model zoo. In response to this potential overlap, we establish the victim QNN on the cloud and also integrate it into the local model zoo alongside other SOTA QNNs. We present comprehensive results detailing the final performance across all possible substitute QNN selections.

\vspace{2pt}\noindent\textbf{VQC Ansatz/Layers \& Ensemble Committee Number}. 
Distinct VQC ansatzs represent different function families and are crucial for the accuracy of a QNN. In this work, we focus on a homogeneous ensemble, where all committee classifiers are implemented with the same base VQC ansatz. This choice aims to mitigate training challenges on NISQ devices. The exploration of heterogeneous VQC ansatzes in an ensemble is deferred to future work.
The inclusion of repeated VQC layers, $L$, in a QNN typically enhances model accuracy, given an ample amount of training data. Similarly, increasing the ensemble committee number, $N_C$, generally contributes to an overall improvement in the accuracy of the ensemble. 
To unveil the full potential of~\design, we instantiate it with an optimal configuration and conduct an exploration of the design space. This exploration involves varying the number of VQC layers and the ensemble committee members to assess their respective impacts on performance and efficacy.

\vspace{-6pt}\subsection{Querying the Victim QNN and Bagging} % https://arxiv.org/pdf/2104.02395.pdf
In alignment with classical model stealing techniques~\cite{tramer2016stealing, papernot:2017practical, Jacson:IJCNN2018, Yu:NDSS2020}, \design~queries the victim QNN on a cloud-based NISQ machine. This involves constructing an unlabeled dataset incorporating publicly available data related to the specific problem domain. Initial data samples, denoted as $S$, are sent to the victim VQC, $Q_V$, on the NISQ machine, producing predicted labels, $Q_V(S)$ (i.e., raw probability output vectors), and thereby forming a training dataset $D=\{S, Q_V(S)\}$. Acknowledging the noise inherent in NISQ computers, \design strategically queries the victim VQC on QNNaaS across multiple rounds, evenly distributing queries throughout the day at intervals of $24/m$ hours, allowing for $m$ (i.e., 3) rounds of queries. Subsequently, the proposed bagging algorithm in Algorithm~\ref{a:bagging_algo} is employed to bootstrap $D$ and construct the sub-dataset.
\design utilizes the complete bootstrap sample for model generation, employing the respective out-of-bag instances for validation. Specifically, each bootstrap sub-dataset comprises $N_Q/N_C$ data samples, randomly selected with replacement from the initial dataset $D$ in each iteration of the bagging process. Maintaining the total query number, $N_Q$, is considered optimal.

\setlength{\textfloatsep}{2pt}
\newcommand\mycommfont[1]{\footnotesize\ttfamily\textcolor{blue}{#1}}
\SetCommentSty{mycommfont}
\SetKwInput{KwInput}{Input}                % Set the Input
\SetKwInput{KwOutput}{Output}              % set the Output
\SetKwRepeat{Do}{do}{while}%
\begin{algorithm}[!t]
\footnotesize
\DontPrintSemicolon
\KwInput{query budget $N_Q$, training dataset $D$, number of bootstrap samples $N_C$, VQC ensemble $Q$,}
\KwOutput{updated ensemble $Q$}

\SetKwProg{Fn}{Procedure}{:}{}
\Fn{Train ($D$, $N_Q$, $N_C$, $Q$)}
{
\For{$i=0$; $i<N_C$; $i++$}
	{ $D_i$ = bootstrap $N_Q$/$N_C$ data from $D$ (i.i.d. sample with replacement)
 
      $D_{oob}$ = out of bag data

      $Q_{i}$ = base classifer trained with $D_i$ 

      $Acc_{i}$ = validation accuracy of $Q_i$ on $D_{oob}$ 

	\If{$Acc_{i} \geq Acc_{i-1}$}
	   {
            train base classifier $Q_i$ with $D_i$
	     }                  
	}
}
\caption{Pseudo code of the bagging algorithm.}
\label{a:bagging_algo}
\vspace{-4pt}
\end{algorithm}

\subsection{Training the Local Substitute Model}
To ensure the robust training of local substitute QNNs, we investigate different loss functions within the \design framework. In addition to the default negative log-likelihood (NLL) loss, we introduce the Huber loss~\cite{Meyer:CVPR2021}. The NLL loss is easily applied through a \texttt{LogSoftmax} function for post-processing on the results obtained from the QNNaaS server. Conversely, the Huber loss, recognized for its robustness, is employed to construct a precise VQC while addressing the challenges posed by noisy labels. The general form of the Huber loss can be defined as follows:
\begin{equation}
L_\delta(a)= \begin{cases}\frac{1}{2} a^2 & \text { for }|a| \leq \delta; \\ \delta \cdot(|a|-\frac{1}{2}\delta), & \text{otherwise.}\end{cases}
\end{equation}
where $a$ is the input; $L_\delta$ indicates the output; and $\delta$ means a predefined value. The Huber loss function is quadratic for small values of $a$, and linear for large values of $a$, with equal values and slopes of the different sections at the two points where $|a|=\delta$. The conventional squared loss $L_s(a)=a^2$ serves as an arithmetic mean-unbiased estimator, but it has the tendency to be dominated by outliers. On the contrary, the Huber loss combines sensitivity of the squared loss, and the robustness of the absolute loss $L_a(a)=|a|$, i.e., a median-unbiased estimator.

\noindent\textbf{Decision Fusion Strategy}.
In our exploration of decision-making strategies within an ensemble, we considered various fusion schemes such as unweighted averaging, majority vote, and weighted methods based on individual influence. Following careful evaluation in the~\design~ framework, we found no discernible differences among candidate methods. As a result, we have chosen the majority vote as the default fusion strategy for its simplicity of implementation.

\section{Potential Defenses against~\design}
\label{s:defenses}
Utilizing watermarking techniques~\cite{Jia:Security2021}, a QNN owner can establish ownership and detect potential theft of their model. Watermarking involves inducing overfitting in the QNN by training it with outlier input-output pairs known only to the defender. Subsequently, these watermarks can be employed to validate ownership of the QNN. The outliers are strategically generated by introducing a unique trigger, such as a small square in a non-intrusive location within an input data sample, effectively serving as distinctive watermarks.  If encountering a model displaying the rare and unexpected behavior encoded by these watermarks, the defender can reasonably infer that the model is a stolen replica.

The deployment of Physical Unclonable Functions (PUF)\cite{Arapinis2021QPUF} and the incorporation of obsolete circuit design through the insertion of dummy \texttt{SWAP} gates\cite{Aakarshitha2021QuantumCircuit} can also potentially contribute to fortifying defenses against information leaks.

\begin{table*}[t!]
\centering
\caption{Comparison of accuracies on substitute QNNs across different model extraction schemes.} 
\vspace{-8pt}
\begin{tabular}{|c||c|c|c|c|c|c|c|}\hline
\multirow{2}{*}{\textbf{Dataset}}  &\multirow{2}{*}{\textbf{Task}} &\multicolumn{2}{c|}{Victim}  & \multicolumn{2}{c|}{CloudLeak} &\multirow{2}{*}{\textbf{QuantumLeak}} & $\uparrow$ over \\\cline{3-6} 
& & Ideal &Real & Ideal &Real & &CloudLeak\_real \\\hline

\multirow{2}{*}{MNIST}   & \multicolumn{1}{l|}{0/1 Classification} &89.55 &86.18 & 85.87 &79.40  &86.75  &\textbf{7.35}    \\ \cline{2-8}
                         & \multicolumn{1}{l|}{2/3 Classification} &84.22 &81.44 &81.01 &76.5  &81.49  &\textbf{4.99}   \\ \hline
\multirow{2}{*}{F-MNIST} & \multicolumn{1}{l|}{t-shirt/trouser} &77.02 &75.47 &73.4 &70.1  &75.67  &\textbf{5.57}   \\ \cline{2-8}
                         & \multicolumn{1}{l|}{pullover/dress} &76.92 &76.05 &72.5 &69.2  &75.84  &\textbf{6.64}    \\ \hline
\end{tabular}
\label{t:result_overall}
\vspace{-0.2in}
\end{table*}

\section{Experimental Methodology}
\label{s:em}

\noindent\textbf{Datasets}. We evaluated~design using the MNIST and Fashion-MNIST datasets. To reduce image dimensions, we applied principal component analysis and average pooling, resulting in $1\times8$ size images.
Similar to previous QNNs~\cite{chen2020variational, Chu:ISLPED2022}, we down-sampled images in MNIST and Fashion-MNIST (i.e., F-MNIST) datasets to the dimensions of $1\times 8$ using principal component analysis and average pooling. 
For MNIST, we focused on the 2-category classifications ($0\&1$ and $2\&3$), while for Fashion-MNIST, we focused on the 2-category classifications (t-shirt/trouser and pullover/dress).

\vspace{2pt}\noindent\textbf{NISQ Machines \& Quantum Compiler}. 
All our experimental designs were executed and assessed using IBM Quantum System backends, specifically harnessing the capabilities of the 27-qubit \texttt{IBM\_Auckland}. To construct the quantum circuits, we utilized BQSKit~\cite{patel2022quest}, employing its default settings for compilation. The deployment of various QNNs on NISQ computers was achieved using Qiskit~\cite{Qiskit}.

\vspace{2pt}\noindent\textbf{Victim VQC \& Training Configuration}. 
The VQC design utilized for the victim QNN) on the QNNaaS cloud server was meticulously crafted through exploration. This optimal VQC consists of a 4-qubit circuit, incorporating an amplitude encoding layer, two parameterized layers, and a measurement layer. Each VQC block consisted of an \texttt{RZ} layer, an \texttt{RY} layer, an additional \texttt{RZ} layer, and a 2-Qubit \texttt{CRX} layer arranged in a ring configuration. During training, the victim VQC utilized either the MNIST or Fashion-MNIST datasets, with the choice of the NLL or Huber loss function. The training process incorporated the Adam optimizer with a learning rate of 1e-3 and weight decay of 1e-4. A batch size of 32 was employed, and the training proceeded for 100 epochs.

\vspace{2pt}\noindent\textbf{The Unlabeled Dataset Construction}. 
For the MNIST dataset, a maximum of 3000 images were randomly sampled from the numerical range of 0$\sim$3 to create an initial dataset for querying the QNNaaS server, representing the images within the defined problem domain. Additionally, 500 images were randomly selected from the corresponding classes for inference purposes. In the case of Fashion-MNIST, a maximum of 3000 images were randomly chosen from t-shirt, trouser, pullover, or dress classes to create an initial dataset for querying the QNNaaS server, serving as the images within the specified problem domain. Furthermore, 500 images were randomly selected from these corresponding classes to constitute the test dataset.

\vspace{2pt}\noindent\textbf{QNN Zoo}.
To effectively mimic the function of the victim VQC, we employed a local QNN zoo. We considered three circuit architectures denoted as follows:
\begin{itemize}[leftmargin=*, nosep, topsep=0pt, partopsep=0pt]
\item $L1$/$L2$/$L3$: 
We incorporate the victim QNN's VQC into the QNN zoo, providing three configurations, each with 1$\sim$3 repeated VQC layers denoted as $L1$/$L2$/$L3$. We assess the entanglement capacity of parameterized VQC using the Meyer-Wallach entanglement metric~\cite{Meyer:JMP2002}. The entanglement capacities for $L1$/$L2$/$L3$ are 0.857, 0.921, and 0.985, respectively.

\item $A1$~\cite{chen2020variational}: $A1$ features two parameterized blocks, each of which comprises 12 one-qubit gates and 4 two-qubit gates. Notably, the entanglement capability of $A1$ is 0.938.

\item $A2$~\cite{Chu:ISLPED2022}: $A2$ consists of two parameterized blocks, which of which has 4 one-qubit gates and 4 two-qubit gates. The entanglement capability of $A2$ is 0.926.
\end{itemize}
We present results regarding the influence of the local substitute QNN on the final model extraction accuracy. For evaluation purposes, we chose $L1$/$L2$/$L3$ as the principal circuit architectures. An Adam optimizer with a learning rate of 1e-3, a weight decay of 1e-4, and either a Negative Log Likelihood (NLL) or Huber loss function was employed to train all local QNNs. The training of local QNNs was executed over 100 epochs.

\vspace{2pt}\noindent\textbf{Attack Schemes}.
To showcase the effectiveness of the techniques proposed in this work, we devised various schemes to assess their impact. These schemes can be classified into the following four categories:
\begin{itemize}[leftmargin=*, nosep, topsep=0pt, partopsep=0pt]
\item \textit{Single-N}: A singular local substitute QNN utilizing NLL loss and integrating the CloudLeak attack.
\item \textit{Single-H}: A singular local substitute QNN utilizing Huber loss and incorporating the CloudLeak attack.
\item \textit{Ens-N}: A QNN ensemble utilizing NLL loss and integrating the~\design~attack.
\item \textit{Ens-H}: A QNN ensemble utilizing Huber loss and incorporating the ~\design~attack.
\end{itemize}

% \vspace{2pt}\noindent\textbf{Evaluation metrics}. 
% In assessing the effectiveness of the quantum model extraction attack, we utilized the inference accuracy of the local substitute QNN on the respective dataset as our primary evaluation metric. A higher accuracy signifies greater similarity between the extracted substitute model and the targeted victim model. It is essential to note that the accuracy of the ideal victim QNN serves as the upper bound in this evaluation process, providing a benchmark against which the success of the extraction attack can be measured.

\vspace{2pt}\noindent\textbf{Design Overhead Comparison}. 
We conducted a comparison of the design overhead between CloudLeak and \design, as presented in Table~\ref{tab:quantum_design_overhead}. Both stealing schemes utilized 6000 images from the problem domain, while CloudLeak also required 512 images from outside the problem domain. For evaluation purposes, we assumed that both schemes performed three rounds of queries to the victim VQC on the cloud-based NISQ machine. Furthermore, the local VQCs trained by both schemes used $L2$ as the its circuit architecture.

\begin{table}[t]
\centering
\vspace{-0.05in}
\caption{The design overhead of CloudLeak and \design.}
\vspace{-0.1in}
\setlength{\tabcolsep}{3pt}
\begin{tabular}{|c||c|c|c|}\hline
            & unlabeled dataset           & query        & local VQC \\
						& size (domain, non-domain)   & round \#     & architecture \\\hline\hline
CloudLeak   & 6000, 512                    & 3            &   $L2$             \\\hline
\design     & 6000, 0                      & 3            &   $L2$              \\
\hline
\end{tabular}
\label{tab:quantum_design_overhead}
\vspace{-0in}
\end{table}

\section{Evaluation and Results}
\subsection{Overall Stealing Accuracy}
\noindent\textbf{Comparison between~\design~\& CloudLeak}.
Table~\ref{t:result_overall} presents a comparative analysis of accuracy between~\design~and CloudLeak. \design~outperforms CloudLeak across all tasks, showcasing an up to 7.35\% accuracy improvement, even when considering NISQ real device noises. Notably,\design~demonstrates comparable accuracy to the ideal victim QNN in specific tasks. While these results may be influenced by the relatively small test dataset (i.e., 500), they underscore the efficacy of~\design~attack, even in the presence of quantum noises.

\begin{figure}[t!]
%\vspace{-0.1in}
\centering
\includegraphics[width=\linewidth]{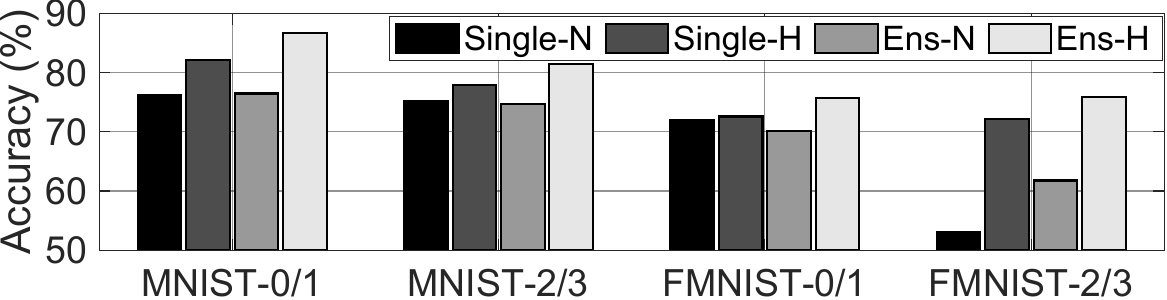}
\vspace{-0.3in}
\caption{The accuracy of~\design~attack on different tasks.}
\label{f:6000-L3-DifferentTasks-v2}
\vspace{-0in}
\end{figure}

\noindent\textbf{Results on Various~\design~Configuration}.
Figure~\ref{f:6000-L3-DifferentTasks-v2} provides a comprehensive accuracy comparison across various tasks. To unlock the full potential of~\design, we first initialize the local substitute QNN as $L2$, maintaining an identical architecture to the cloud victim QNN.
Our experiments encompass four schemes across four distinct tasks: \textit{MNIST-0/1} and \textit{MNIST-2/3} for MNIST 2-category classifications, and \textit{FNIST-0/1} and \textit{FMNIST-2/3} for Fashion-MNIST 2-category classifications. Utilizing a maximum query budget of 6,000, the results yield the following insights: 
(1) The ensemble of QNNs consistently outperforms singular QNNs across most scenarios, except for \textit{MNIST-0/1} and \textit{FMNIST-0/1}, where \textit{Ens-N} exhibits a 0.39\% and 0.18\% accuracy reduction, respectively, compared to \textit{Single-N}.
(2) The application of Huber loss consistently enhances training effectiveness in all cases, demonstrating its efficacy in learning from noisy labels.
(3) \textit{Ens-H}, the combination of QNN ensemble and Huber loss, achieves the highest performance across all tasks, underscoring its superiority in achieving optimal outcomes. In comparison with the baseline accuracy of CloudLeak, as shown in Figure~\ref{f:motivation} (i.e., 79.4\%, 76.5\%, 70.1\%, 69.2\%), \textit{Ens-H} achieves respective accuracy improvements of 7.35\%, 4.99\%, 5.57\%, and 6.64\%.
\textit{These results not only showcase the effectiveness of the~\design~framework but also emphasize the crucial role played by the quantum ensemble and robust Huber loss in its implementation.}

\subsection{Impact of VQC Ansatz}
\begin{figure}[t!]
\centering
\includegraphics[width=\linewidth]{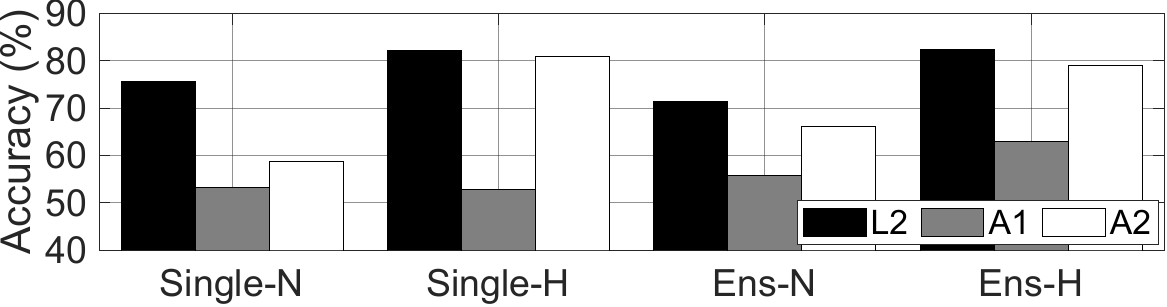}
\vspace{-0.3in}
\caption{The accuracy of QNNs with different VQCs.}
\label{f:AnsatzTypeV2}
\vspace{-0in}
\end{figure}

We explore the influence of substitute model structures on the performance of the~\design~attack and report the results in Figure~\ref{f:AnsatzTypeV2}. All three QNNs incorporate dense angle encoding with increased encoding density and exhibit similar entanglement capabilities. it is not surprising that $L2$ achieves the best accuracy as it is hand-crafted to for~\design~.
However, a noticeable accuracy decline characterizes $A1$ in contrast to both $L2$ and $A2$. This decline is predominantly ascribed to the implementation of fixed maximum entanglement 2-Qubit \texttt{CNOT} gates in $A1$. As indicated by prior research~\cite{Chu:ISLPED2022}, the ability to dynamically adapt and learn task-specific entanglement through parameterized two-qubit gates is pivotal for enhancing QNN accuracy.
\textit{These findings underscore the significant impact of the choice of model structure on the efficacy of the \design~attack. Locally well-crafted substitutes, exemplified by $L2$, consistently outperform other candidates. The observed decrease in accuracy for $A1$ further highlights the imperative of prioritizing task-specific entanglement over entanglement capacity as a valid metric when selecting the local substitute QNN.}

\subsection{Impact of Ensemble Committee Number}
\begin{figure}[t!]
\centering
\includegraphics[width=\linewidth]{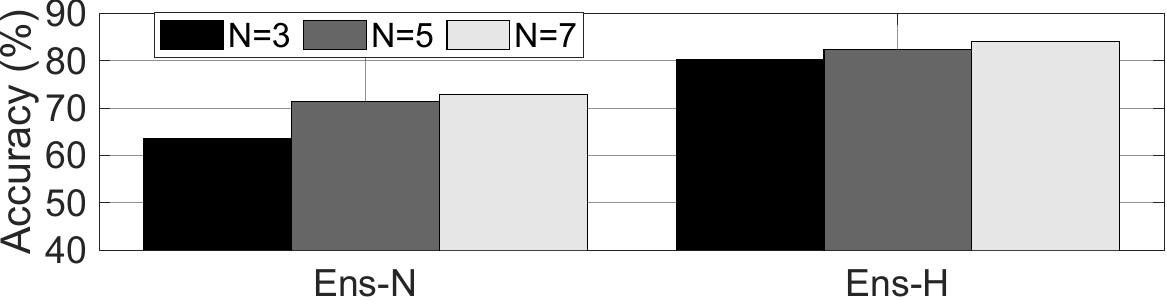}
\vspace{-0.3in}
\caption{The accuracy of~\design~attacks with various committee number.}
\label{f:ClassifierNumberV2}
\vspace{-0in}
\end{figure}

We investigate the impact of varying ensemble committee numbers on the performance of the \design~attack and report the accuracy results in Figure~\ref{f:ClassifierNumberV2}. Throughout the experimentation, the local substitute QNN backbone remains constant with the utilization of $L2$. The results consistently demonstrate a monotonically increasing trend in accuracy as the committee number increases.
Specifically, augmenting the committee number from 3 to 5 results in a 7.67\% and 2.5\% improvement in accuracy for the \texttt{Ens-H} and \texttt{Ens-H} schemes, respectively. Contrarily, further increasing the committee number from 5 to 7 yields a 1.59\% and 1.4\% accuracy improvement for the \texttt{Ens-H} and \texttt{Ens-H} schemes. \textit{These results underscore the efficacy of increasing the committee number in enhancing the attack performance of \design. However, it is crucial to note the observed performance gains diminish, leading to a practical trade-off between improved performance and the associated training and hardware overhead.}

\vspace{-4pt}
\subsection{Ablation Study on Query Budget and VQC Layers}
\begin{figure*}[t!]
\centering
\includegraphics[width=\textwidth]{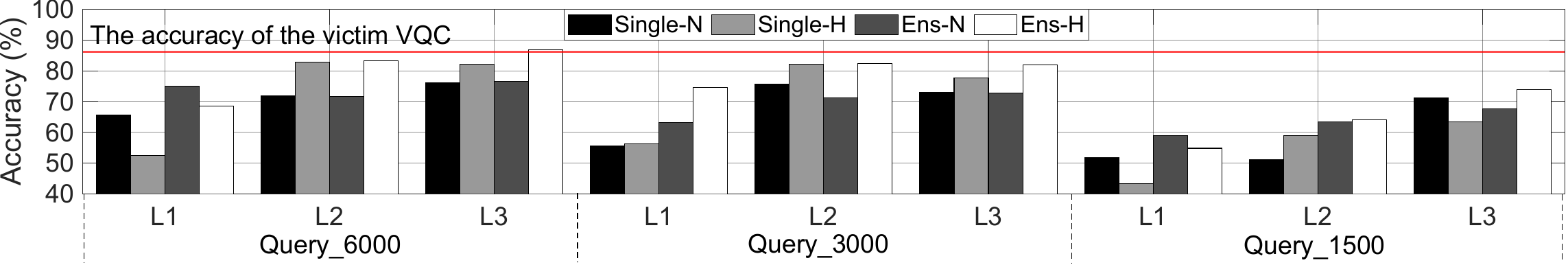}
\vspace{-0.3in}
\caption{Performance of local VQCs with different layers.}
\label{f:LayerComparisonV2}
\vspace{-0.1in}
\end{figure*}

\vspace{-4pt}
In our investigation into the impact of query number budget on the effectiveness of~\design~ attack, we conducted an ablation study by systematically varying the query number for QNNaaS to 1,500, 3,000, and 6,000, respectively. The resulting accuracies of a locally deployed substitute QNN, featuring architectures $L1$, $L2$, and $L3$, are presented in Figure~\ref{f:LayerComparisonV2}.
Our findings highlight a correlation between the augmentation of the query budget and the acquired accuracy of the local QNN. Noteworthy is the observation that the $L3$ architecture, leveraging 6,000 queries, attains an accuracy comparable to that of the victim QNN, signifying the successful replication of QNNaaS functionality. However, it is important to underscore that an escalated query count heightens the susceptibility of the local adversary to detection, thereby compromising the stealthiness of the model extraction process.
Moreover, we observed that, for a given query budget, the augmentation of QNN ansatz layers contributes to an enhancement in final accuracy. \textit{This introduces a nuanced trade-off between the efficacy of the design attack and the associated hardware cost. It becomes paramount, therefore, to consider and strike a balance between these factors in the pursuit of an optimal solution, emphasizing the importance of a judicious approach to the deployment of resources in adversarial scenarios.}

\begin{figure*}[t!]
\centering
\includegraphics[width=\textwidth]{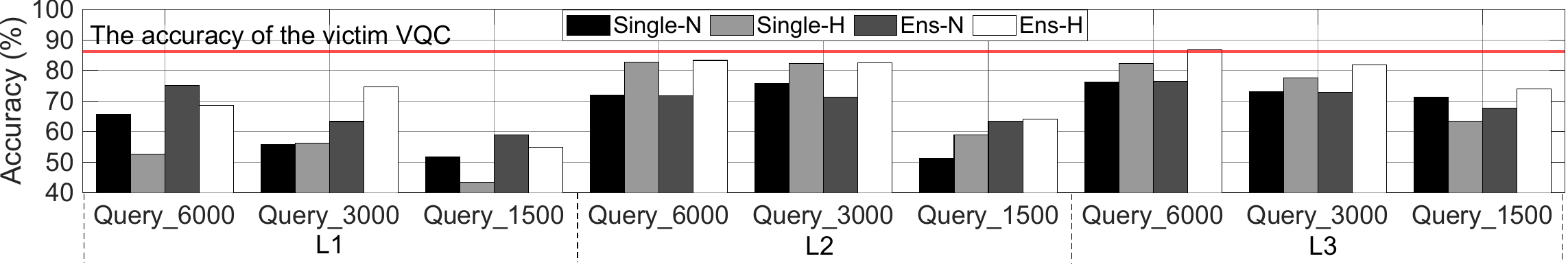}
\vspace{-0.3in}
\caption{Performance of local VQCs with different query times.}
\label{f:QueryTimeComparisonV2}
\vspace{-0.26in}
\end{figure*}

Figure~\ref{f:QueryTimeComparisonV2} provides the accuracy results for a given QNN architecture across various query budgets. Notably, $L3$ exhibits a pronounced accuracy increase as the query budget is escalated. For instance, the accuracy of \textit{Ens-H} on $L3$ rose from 73.93\% to 86.75\% with a query budget increase from 1,500 to 6,000. Contrastingly, the results for $L1$ reveal a counterintuitive trend. The accuracy of \textit{Ens-H} on $L1$ at query budgets of 1,500, 3,000, and 6,000 is 54.8\%, 74.62\%, and 68.56\%, respectively. Notably, there is a significant 6.06\% reduction when the query number is further increased from 3,000 to 6,000. This phenomenon can be primarily attributed to the overfitting of the $L1$ QNN model. Therefore, when operating within a design-based framework, especially when employing a simple local substitution model, it is imperative to validate whether the model is prone to overfitting. \textit{Large query budgets may inadvertently decrease the performance of the final extracted local model, emphasizing the critical need for careful model validation and optimization in~\design~attacks.}

\section{Conclusion}
In conclusion, \design introduces a novel model extraction attack for acquiring Quantum Neural Networks (QNNs) from cloud-based NISQ computers. By leveraging an ensemble of local QNNs and capitalizing on their collective decision-making capabilities, \design significantly enhances the accuracy of the local substitute QNN. To further fortify the robustness of local substitute QNN training with queried noisy data, we propose a generalized formulation of the Huber loss within the \design framework. Experimental results on IBM NISQ devices demonstrate that \design achieves a notable improvement of 4.99\%$\sim$7.35\% in the accuracy of the locally-generated QNN compared to the state-of-the-art classical model extraction attack.

\section*{Acknowledgment}
This work was supported in part by NSF CAREER AWARD CNS-2143120. 
We thank the IBM Quantum Researcher \& Educators Program for their support of Quantum Credits.

\vspace{-6pt}
\bibliographystyle{ieeetr}
\bibliography{reference}

\begin{thebibliography}{10}

\bibitem{chen2020variational}
S.~Chen, C.~Yang, J.~Qi, P.~Chen, X.~Ma, and H.~Goan, ``{Variational Quantum Circuits for Deep Reinforcement Learning},'' {\em IEEE Access}, vol.~8, pp.~141007--141024, 2020.

\bibitem{Chu:ISLPED2022}
C.~Chu, N.-H. Chia, L.~Jiang, and F.~Chen, ``{QMLP: An Error-Tolerant Nonlinear Quantum MLP Architecture Using Parameterized Two-Qubit Gates},'' in {\em ACM/IEEE International Symposium on Low Power Electronics and Design (ISLPED)}, 2022.

\bibitem{Egger:TQE2020}
D.~Egger, C.~Gambella, J.~Marecek, S.~McFaddin, M.~Mevissen, R.~Raymond, A.~Simonetto, S.~Woerner, and E.~Yndurain, ``{Quantum Computing for Finance: State-of-the-Art and Future Prospects},'' {\em IEEE Transactions on Quantum Engineering}, 2020.

\bibitem{Mitarai:physA2018}
K.~Mitarai, M.~Negoro, M.~Kitagawa, and K.~Fujii, ``{Quantum Circuit Learning},'' {\em Phys. Rev. A}, 2018.

\bibitem{tramer2016stealing}
F.~Tram{\`e}r, F.~Zhang, A.~Juels, M.~K. Reiter, and T.~Ristenpart, ``{Stealing Machine Learning Models via Prediction APIs},'' in {\em USENIX security symposium}, vol.~16, 2016.

\bibitem{papernot:2017practical}
N.~Papernot, P.~McDaniel, I.~Goodfellow, S.~Jha, Z.~B. Celik, and A.~Swami, ``{Practical Black-Box Attacks against Machine Learning},'' in {\em ACM on Asia Conference on Computer and Communications Security}, pp.~506--519, 2017.

\bibitem{Jacson:IJCNN2018}
J.~R. {Correia-Silva}, R.~F. {Berriel}, C.~{Badue}, A.~F. {de Souza}, and T.~{Oliveira-Santos}, ``{Copycat CNN: Stealing Knowledge by Persuading Confession with Random Non-Labeled Data},'' in {\em International Joint Conference on Neural Networks}, 2018.

\bibitem{Yu:NDSS2020}
H.~Yu, K.~Yang, T.~Zhang, Y.~Tsai, T.-Y. Ho, and Y.~Jin, ``{CloudLeak: Large-Scale Deep Learning Models Stealing Through Adversarial Examples},'' in {\em Annual Network and Distributed System Security Symposium (NDSS)}, 2020.

\bibitem{Clerk:RMP2010}
A.~A. Clerk, M.~H. Devoret, S.~M. Girvin, F.~Marquardt, and R.~J. Schoelkopf, ``{Introduction to Quantum Noise, Measurement, and Amplification},'' {\em Reviews of Modern Physics}, vol.~82, Apr 2010.

\bibitem{sarovar2020detecting}
M.~Sarovar, T.~Proctor, K.~Rudinger, K.~Young, E.~Nielsen, and R.~Blume-Kohout, ``{Detecting Crosstalk Errors in Quantum Information Processors},'' {\em Quantum}, vol.~4, p.~321, 2020.

\bibitem{Huo:NJP2017}
M.-X. Huo and Y.~Li, ``{Learning Time-Dependent Noise to Reduce Logical Errors: Real-Time Error Rate Estimation in Quantum Error Correction},'' {\em New Journal of Physics}, vol.~19, no.~12, p.~123032, 2017.

\bibitem{Lax:PR1966}
M.~Lax, ``{Quantum Noise. IV. Quantum Theory of Noise Sources},'' {\em Physical Review}, vol.~145, no.~1, p.~110, 1966.

\bibitem{Zhu:SCIENCE2019}
D.~Zhu, N.~M. Linke, M.~Benedetti, K.~A. Landsman, N.~H. Nguyen, C.~H. Alderete, A.~Perdomo-Ortiz, N.~Korda, A.~Garfoot, C.~Brecque, L.~Egan, O.~Perdomo, and C.~Monroe, ``{Training of Quantum Circuits on a Hybrid Quantum Computer},'' {\em Science Advances}, vol.~5, no.~10, 2019.

\bibitem{ryan2021realization}
C.~Ryan-Anderson, J.~G. Bohnet, K.~Lee, D.~Gresh, A.~Hankin, J.~Gaebler, D.~Francois, A.~Chernoguzov, D.~Lucchetti, N.~C. Brown, {\em et~al.}, ``{Realization of Real-Time Fault-Tolerant Quantum Error Correction},'' {\em Physical Review X}, vol.~11, no.~4, p.~041058, 2021.

\bibitem{nielsen2002quantum}
M.~A. Nielsen and I.~L. Chuang, {\em {Quantum Computation and Quantum Information}}.
\newblock Cambridge University Press, 2010.

\bibitem{rudinger2019probing}
K.~Rudinger {\em et~al.}, ``{Probing Context-Dependent Errors in Quantum Processors},'' {\em Physical Review X}, 2019.

\bibitem{ahsan2022quantum}
M.~Ahsan, S.~A.~Z. Naqvi, and H.~Anwer, ``{Quantum Circuit Engineering for Correcting Coherent Noise},'' {\em Physical Review A}, vol.~105, no.~2, p.~022428, 2022.

\bibitem{nachman2020unfolding}
B.~Nachman, M.~Urbanek, W.~A. de~Jong, and C.~W. Bauer, ``{Unfolding Quantum Computer Readout Noise},'' {\em npj Quantum Information}, vol.~6, no.~1, p.~84, 2020.

\bibitem{Sukin2019_vqcc14}
S.~Sim, P.~D. Johnson, and A.~Aspuru-Guzik, ``{Expressibility and Entangling Capability of Parameterized Quantum Circuits for Hybrid Quantum-Classical Algorithms},'' {\em Advanced Quantum Technologies}, vol.~2, no.~12, p.~1900070, 2019.

\bibitem{hansen:1990neural}
L.~Hansen and P.~Salamon, ``{Neural Network Ensembles},'' {\em IEEE transactions on pattern analysis and machine intelligence}, vol.~12, no.~10, pp.~993--1001, 1990.

\bibitem{ferreira2012boosting}
A.~J. Ferreira and M.~A. Figueiredo, ``{Boosting Algorithms: A Review of Methods, Theory, and Applications},'' {\em Ensemble machine learning: Methods and applications}, pp.~35--85, 2012.

\bibitem{Meyer:CVPR2021}
G.~P. Meyer, ``{An Alternative Probabilistic Interpretation of the Huber Loss},'' in {\em IEEE/CVF Conference on Computer Vision and Pattern Recognition}, pp.~5261--5269, 2021.

\bibitem{Jia:Security2021}
H.~Jia, C.~A. Choquette-Choo, V.~Chandrasekaran, and N.~Papernot, ``{Entangled Watermarks as a Defense Against Model Extraction},'' in {\em USENIX Security Symposium}, pp.~1937--1954, 2021.

\bibitem{Arapinis2021QPUF}
M.~Arapinis, M.~Delavar, M.~Doosti, and E.~Kashefi, ``{Quantum Physical Unclonable Functions: Possibilities and Impossibilities},'' {\em Quantum}, vol.~5, p.~475, 2021.

\bibitem{Aakarshitha2021QuantumCircuit}
A.~Suresh, A.~Ash{-}Saki, M.~Alam, R.~O. Topaloglu, and S.~Ghosh, ``{{A} Quantum Circuit Obfuscation Methodology for Security and Privacy},'' in {\em Workshop on Hardware and Architectural Support for Security and Privacy, Virtual Event}, 2021.

\bibitem{patel2022quest}
T.~Patel, E.~Younis, C.~Iancu, W.~de~Jong, and D.~Tiwari, ``{Quest: Systematically Approximating Quantum Circuits for Higher Output Fidelity},'' in {\em ACM International Conference on Architectural Support for Programming Languages and Operating Systems (ASPLOS)}, pp.~514--528, 2022.

\bibitem{Qiskit}
{Qiskit contributors}, ``{Qiskit: An Open-source Framework for Quantum Computing},'' 2023.

\bibitem{Meyer:JMP2002}
D.~A. Meyer and N.~R. Wallach, ``{Global Entanglement in Multiparticle Systems},'' {\em Journal of Mathematical Physics}, vol.~43, no.~9, pp.~4273--4278, 2002.

\end{thebibliography}

\end{document}